\documentclass{PoS}
\usepackage{amssymb,amsmath}

\title{NSPT calculations in the Schr\"odinger Functional formalism}

\ShortTitle{NSPT calculations in the SF formalism}

\author{\speaker{C.~Torrero}, G.S. Bali\\
        Institute for Theoretical Physics, University of Regensburg, 93040, Regensburg, Germany\\
        E-mail: \email{christian.torrero@physik.uni-regensburg.de}\\
        \hspace*{1.2cm}\email{gunnar.bali@physik.uni-regensburg.de}}

\abstract{Within the framework of the Schr\"odinger Functional (SF), we outline how to combine Numerical Stochastic Perturbation Theory (NSPT) and PCAC relations to determine the two-loop contributions to the improvement coefficients $c_A$ and $c_{SW}$ for Sheikholeslami-Wohlert-Wilson fermions.}

\FullConference{The XXVII International Symposium on Lattice Field Theory - LAT2009\\
		 July 26-31 2009\\
		 Peking University, Beijing, China}

\begin{document}

\section{Introduction}

As it is well-known, in the improvement approach \`a la Symanzik~\cite{Sym} the lattice QCD action has to be provided with an extra irrelevant contribution, the so-called \emph{Sheikholeslami-Wohlert term}~\cite{CSW}. In Perturbation Theory (PT), it features a scalar coefficient $c_{SW}$ which can be Taylor-expanded in powers of the bare coupling $g_0$ as

\vspace*{-0.1cm}
\begin{equation}
c_{SW} = c_{SW}^{(0)}\ \! +\ \! c_{SW}^{(1)}g_0^2\ \! +\ \! c_{SW}^{(2)}g_0^4\ \! +\ \! \mathcal{O}(g_0^6)\ .
\end{equation}
\vspace*{-0.25cm}

\hspace*{-0.7cm}The zero- and one-loop coefficients have already been determined for different lattice actions~\cite{Woh}\cite{QCDSF} while $c_{SW}^{(2)}$ is still unknown: the final aim of this project is precisely to estimate it by combining the \emph{Schr\"odinger Functional formalism} (SF) and the \emph{PCAC relations} in the same spirit as \cite{LW1} and \cite{LW2} where $c_{SW}^{(0)}$ and $c_{SW}^{(1)}$ were successfully recovered.\\
\hspace*{0.65cm} The main difference with these two latter seminal papers lies in the fact that observables are evaluated perturbatively without following a diagrammatic approach but rather by means of \emph{Numerical Stochastic Perturbation Theory} (NSPT), a computer algorithm characterized by a Langevin-like evolution of the system.


\section{Theoretical aspects - part I (basics)}

The lattice formulation of QCD we adopt is that of Wilson: a concrete expression of the well-known contributions to the action - namely the gauge ($S_G)$, fermionic ($S_F$) and Sheikholeslami-Wohlert ($S_{SW}$) term - can be found in \cite{LW1} whose notations and conventions inspire nearly all the formulae appearing in this and the next section$~\!$\footnote{$~\!$More generally, we stick to the setup outlined in sections 2, 4 and 6 of \cite{LW1}.}.

A suitable observable to study in order to evaluate $c_{SW}^{(2)}$ is provided by the quark mass $m_q$ which can be conveniently computed by means of the lattice PCAC relation reading,$~\!$\footnote{$~\!$From now on, the time direction will be assigned the subscript 0.}

\begin{equation}
\frac{1}{2}(\partial_0^R+\partial_0^L)\langle A_0^b(n)\mathbb{O}\rangle=2m_q\langle P^b(n)\mathbb{O}\rangle\ ,
\end{equation}  
\vspace*{-0.2cm}

\hspace*{-0.7cm}where $\mathbb{O}$ is any product of fields located at nonzero distance from $n$, $\partial_0^R$ ($\partial_0^L$) is the lattice right (left) derivative in the time direction and

\begin{equation}
A_0^b(n) = \sum_{f\!,g}^{N_f}\overline{\psi}^f\!(n)\gamma_{\mu}\gamma_5\frac{1}{2}\tau^b_{fg}\psi^g(n)\ , \ \ \ \ \ \ P^b(n) = \sum_{f\!,g}^{N_f}\overline{\psi}^f\!(n)\gamma_5\frac{1}{2}\tau^b_{fg}\psi^g(n)\ ,
\end{equation}
\vspace*{-0.1cm}

\hspace*{-0.7cm}where $\tau^b$ is a matrix acting on flavour degrees of freedom$~\!$\footnote{$~\!$Spin and colour subscripts will be usually left implicit in order to ease the notation.}.\\
\hspace*{0.7cm}In order to fix $c_{SW}^{(2)}$, one requires $m_q$ to be independent of contributions of order $a$: however, to achieve full improvement Eq.(2.1) has to be modified to,

\vspace*{-0.25cm}
\begin{eqnarray}
\frac{1}{2}(\partial_0^R+\partial_0^L)\langle A_0^b(n)\mathbb{O}\rangle + c_A\partial_0^L\partial_0^R\langle P^b(n)\mathbb{O}\rangle&=&2m_q\langle P^b(n)\mathbb{O}\rangle\ ,\\
&&\nonumber
\end{eqnarray}
\vspace*{-0.3cm}

\hspace*{-0.7cm}where $c_a$ is a second improvement coefficient which, just like $c_{SW}$, can also be decomposed as $c_A = c_{A}^{(0)}\ \! +\ \! c_{A}^{(1)}g_0^2\ \! +\ \! c_{A}^{(2)}g_0^4\ \! +\ \! \mathcal{O}(g_0^6)$. Once again, the first unknown contribution is at two-loop level: see \cite{LW1} and \cite{LW2} for the determination of $c_{A}^{(0)}$ and $c_{A}^{(1)}$.\\
\hspace*{0.7cm}The second main theoretical ingredient of the present strategy is given by the Schr\"odinger Functional: assuming the time coordinate ranges from $0$ to $T$ and labelling the space coordinates as $\vec{n}$, it consists of replacing the usual periodic boundaries by Dirichlet conditions along the time direction, namely,

\vspace*{-0.05cm}
\begin{equation}
\left.U_{k}(n)\right|_{n_0=0} \rightarrow\ W_k(\vec{n})\ , \ \ \ \ \ \ \ \ \ \left.U_{k}(n)\right|_{n_0=T} \rightarrow\ W'_k(\vec{n}) \ \ \ \ (k = 1,2,3)\ ,
\end{equation}  
\vspace*{-0.25cm}

\hspace*{-0.7cm}for the gauge degrees of freedom$~\!$\footnote{$~\!$Gauge fields along the time direction, defined for $0\leq n_0<T$, have no constraints on them. It turns out that $W$ and $W'$ can sloppily be written as $W=\mathcal{P}e^{\int C}$ and $W=\mathcal{P}e^{\int C'}$ - see section 6 of \cite{LW1} for notations and a more careful and detailed treatment of this topic - where $C$ and $C'$ play a similar role as the background field in classical physics: in what follows we will refer to the case $C=C'=0$ as the \emph{trivial background}.} and ($P_{\pm}=(\mathbb{I}\pm\gamma_0)/2$ with $\mathbb{I}$ being the identity matrix)

\vspace*{-0.3cm}
\begin{eqnarray}
\!\!\!\!\!\!\!\!\!\!\left.\psi^f\!(n)\right|_{n_0=0}\ \rightarrow\ \rho^f\!(\vec{n})=P_+\!\left.\psi^f\!(n)\right|_{n_0=0}\ , & &\ \ \ \ \  \left.\psi^f\!(n)\right|_{n_0=T}\ \rightarrow\ \rho'^f\!(\vec{n})=P_+\!\left.\psi^f\!(n)\right|_{n_0=T}\ ,\\
\!\!\!\!\!\!\!\!\!\!\left.\overline{\psi}^f\!(n)\right|_{n_0=0}\ \rightarrow\ \overline{\rho}^f\!(\vec{n})=P_+\!\left.\overline{\psi}^f\!(n)\right|_{n_0=0}\ , & &\ \ \ \ \ \left.\overline{\psi}^f\!(n)\right|_{n_0=T}\ \rightarrow\ \overline{\rho}'^f\!(\vec{n})=P_+\!\left.\overline{\psi}^f\!(n)\right|_{n_0=T}\ ,  
\end{eqnarray}
\vspace*{-0.2cm}

\hspace*{-0.7cm}for fermions: boundary fields $W$, $W'$, $\rho$, $\overline{\rho}$, $\rho'$ and $\overline{\rho}'$ will be defined later on.\\
\hspace*{0.7cm}Due to the Schr\"odinger Functional formalism, the three contributions to the lattice QCD action get modified as follows:

\begin{itemize}

\item the gauge part $S_G$ becomes

\vspace*{-0.2cm}
\begin{equation}
S_G = \beta\!\!\!\sum_{\scriptstyle n,~\!\!\mu,~\!\!\!\nu \atop \scriptstyle\mu~\!\!\!>\nu}\omega_{\mu\nu}(n)\bigg(1-\frac{Tr}{2N_c}\big[U_{\mu\nu}(n)+{U}_{\mu\nu}^{\dagger}(n)\big]\bigg)~, \\
\end{equation}
\vspace*{-0.3cm}

where the weight $\omega_{\mu\nu}(n)$ for the lattice plaquette $U_{\mu\nu}(n)\ \!$is $1$ everywhere except for the spatial plaquette at $n_0=0$ and $n_0=T$ whose $\omega_{\mu\nu}(n)$ reads $\frac{1}{2}$;

\item the fermionic part $S_F$ remains in principle unchanged; anyway, in order to have one more parameter to play with, an additional phase $e^{i\theta_{\mu}/L_{\mu}}$ is introduced in the definition of the lattice covariant derivatives within the Wilson-Dirac operator: in practice, gauge fields $U_{\mu}(n)$ appearing in $S_F$ are replaced by,

\vspace*{-0.4cm}
\begin{equation}
U_{\mu}(n) \rightarrow e^{i\theta_{\mu}/L_{\mu}}U_{\mu}(n) \ ,
\end{equation}   

with $\ \!\theta_0 = 0\ \!$ and $\ \!-\pi<\theta_k\leq\pi\ \!$ for $\ \!k=1,2,3$;

\item the clover term is set to 0 for all those lattice points with $n_0=0$ or $n_0=T$.

\end{itemize}


\section{Theoretical aspects - part II (details)}

Before outlining the procedure that should lead to an estimate of $c_{SW}^{(2)}$, let us give a precise shape to the observable $\mathbb{O}$ appearing in Eq.(2.3): a convenient choice reads,

\vspace*{-0.1cm}
\begin{equation}
\mathbb{O} = \ \!a^6\ \!\sum_{f\!,g}^{N_f}\ \!\sum_{\vec{m},\vec{m}'}\overline{\varsigma}^f\!(\vec{m})\gamma_5\frac{1}{2}\tau^b_{fg}\varsigma^g\!(\vec{m}')\ ,
\end{equation}
\vspace*{-0.1cm}

\hspace*{-0.7cm}where

\vspace*{-0.15cm}
\begin{equation}
\varsigma^f\!(\vec{m}) = \frac{\delta}{\delta\overline{\rho}^f\!(\vec{m})}\ , \ \ \ \ \ \ \ \overline{\varsigma}^f\!(\vec{m}) = -\frac{\delta}{\delta\rho^f\!(\vec{m})}\ .
\end{equation}
\vspace*{0.02cm}

After first plugging Eq.(3.1) into Eq.(2.3), then letting the derivatives with respect to $\rho$ and $\overline{\rho}$ act on the Boltzmann factor and finally setting all the fermionic boundary fields to zero, some algebra allows one to write

\vspace*{-0.0cm}
\begin{equation}
m_q = \frac{\frac{1}{2}\big[\frac{1}{2}(\partial_0^R+\partial_0^L)f_A + c_A\partial_0^L\partial_0^Rf_P\big]}{f_P}\ ,
\end{equation} 
\vspace*{-0.1cm}

\hspace*{-0.7cm}with$~\!$\footnote{$~\!$The subscript ``G" stands for the mean over gauge degrees of freedom. Here and in Eqs.(3.6)-(3.7) repeated indices are summed over. Moreover, from now on we tacitly assume that all quantities are rescaled with $a$ to be dimensionless.}

\vspace*{-0.35cm}
\begin{eqnarray}
f_A &=& \frac{1}{12}\sum_{\vec{m},\vec{m}'}\langle H^{lf}_{[(\vec{m}+\hat{0})\ \!\!\omega\ \!\! c\ \!\!,\ \!n\ \!\!\epsilon\ \!\! e]}\ \!(\gamma_0)_{\epsilon\beta}\ \tau^b_{fg}\ \!\big(\!P_-\!\big)_{\!\omega\sigma}\ J^{gh}_{[(\vec{m}'+\hat{0})\ \!\!\sigma\ \!\! c\ \!\!,\ \!n\ \!\!\beta\ \!\! e]}\ \!\tau^b_{hl}\ \!\rangle_{\!\scriptstyle {_G}}\ ,\\
f_P &=& \frac{1}{12}\sum_{\vec{m},\vec{m}'}\langle H^{lf}_{[(\vec{m}+\hat{0})\ \!\!\omega\ \!\! c\ \!\!,\ \!n\ \!\!\epsilon\ \!\! e]}\ \tau^b_{fg}\ \!\big(\!P_-\!\big)_{\!\omega\sigma}\ J^{gh}_{[(\vec{m}'+\hat{0})\ \!\!\sigma\ \!\! c\ \!\!,\ \!n\ \!\!\epsilon\ \!\! e]}\ \!\tau^b_{hl}\ \!\rangle_{\!\scriptstyle {_G}}\ ,\\
&&\nonumber\\
&&\nonumber
\end{eqnarray}
\vspace*{-1.35cm}

\hspace*{-0.7cm}with

\vspace*{-0.55cm}
\begin{eqnarray}
H^{lf}_{[(\vec{m}+\hat{0})\ \!\!\omega\ \!\! c\ \!\!,\ \!n\ \!\!\epsilon\ \!\! e]} &=&  \bigg[U_0(\vec{m})\bigg]_{\!cb}\bigg(\!\widetilde{M}^{-1}\!\bigg)^{\!\!lf}_{\![(\vec{m}+\hat{0})\ \!\! \omega\ \!\! b\ \!\!,\ \!n\ \!\!\epsilon\ \!\! e]}\ \ , \\
J^{gh}_{[(\vec{m}'+\hat{0})\ \!\!\sigma\ \!\! c\ \!\!,\ \!n\ \!\!\beta\ \!\! e]} &=& \bigg[U_0(\vec{m}')\bigg]^{\!*}_{\!cd}\bigg(\! \widetilde{M}^{-1^*}\!\bigg)^{\!\!gh}_{\![(\vec{m}'+\hat{0})\ \!\!\sigma\ \!\! d\ \!\!,\ \!n\ \!\!\beta\ \!\! e]}\ \ ,
\end{eqnarray}
\vspace*{-0.15cm}

\hspace*{-0.7cm}where $\widetilde{M}$ is the overall fermionic opearator in the lattice action.\\
\hspace*{0.7cm}$f_A$, $f_P\ \!$ and $\ \!m_q\ \!$ depend on the lattice spacing $\ \!a$, the lattice extents $\ \!L_{\mu}$, the bare coupling $\ \!g_0$, the gauge fields $\ \!W$ and $\ \!W'$, the angles $\ \!\theta_k$ (from now on, we will set the latter equal to a common value $\theta$) and the improvement coefficients: recalling that the approach is perturbative, we can write$~\!$\footnote{$~\!$We make the dependence on $\ \!W$, $W'\ \!$, $c_{SW}$ and $c_A$ implicit not to overwhelm the notation; at the same time, we drop the subscript on the lattice extents for a reason that will become clear soon.},

\vspace*{-0.4cm}
\begin{eqnarray}
\!\!\!\!\!\!m_q(L,\theta,x_0,g_0,a)\! &=&\! m_q^{(0)}(L,\theta,x_0,a) + m_q^{(2)}(L,\theta,x_0,a)g_0^2 + m_q^{(4)}(L,\theta,x_0,a)g_0^4 + \mathcal{O}(g_0^6)\ ,\\
&&\nonumber
\end{eqnarray} 
\vspace*{-0.4cm}

\hspace*{-0.7cm}and in turn, thanks to dimensional analysis

\vspace*{-0.3cm}
\begin{equation}
m_q^{(k)}(L,\theta,x_0,a) = d_L(c^{(i\le k)}_{SW},c^{(i\le k)}_A)\frac{a}{L} + d_{x_0}(c^{(i\le k)}_{SW},c^{(i\le k)}_A)\frac{a}{x_0} + d_{\theta}(c^{(i\le k)}_{SW},c^{(i\le k)}_A)\frac{a\theta}{L} + \mathcal{O}(a^2)\ .
\end{equation}
\vspace*{-0.1cm}

\hspace*{-0.7cm}This formula can actually be simplified by setting the $L_k$'s to the same value $L$, putting $L_0 = 2L$ and choosing $n_0 = L$: thus, the corrections in $a$ to $m_q^{(k)}$ will be grouped together into a single one proportional to $\ \!a/L$. Since the aim of improvement is to get rid of lattice artifacts of order $\ \!a$, it is reasonable to estimate $\ \!c_{SW}^{(2)}$ by requiring the only coefficient $\ \!d(c^{(i\le k)}_{SW},c^{(i\le k)}_A)$ left in the formula above - after its reduction - to vanish. This can be achieved by the following steps: 1) fix $\ \!c_{SW}^{(2)}$ and $\ \!c_{A}^{(2)}$ arbitrarily after setting $\ \!c_{SW}^{(0)}$, $\ \!c_{SW}^{(1)}$, $\ \!c_{A}^{(0)}$ and $\ \!c_{A}^{(1)}$ to their known values; 2) perform simulations for different lattice extents keeping $\theta$, $W$ and $W'$ constant; 3) fit the coefficient $\ \!d(c_{SW}^{(2)},c_A^{(2)})$; 4) repeat the previous steps for different choices of $\ \!c_{SW}^{(2)}$ and $\ \!c_{A}^{(2)}$; 5) collect the various estimates of $d(c_{SW}^{(2)},c_A^{(2)})$ and interpolate the values of $\ \!c_{SW}^{(2)}$ and $\ \!c_{A}^{(2)}$ for which $d(c_{SW}^{(2)},c_A^{(2)})$ vanishes.







\vspace*{0.5cm}

Before ending this section, some remarks are in order.\\
\hspace*{0.7cm}The first term on the r.h.s. $\!\!\!$of Eq.(3.8) should normally correspond to the bare mass $\widehat{M}_0$ appearing in $S_F$; however, in the present setup, \emph{this is the case only if $\ \!\theta=0$}: we chose to set $\ \!\widehat{M}_0=0\ \!$ but to work with non-vanishing $\theta$ to avoid any infrared divergence.\\
\hspace*{0.7cm}Second, in Eq.(3.9) it is understood that mass counterterms - depending on $c_{SW}$~\cite{PAN} - are subtracted. Otherwise $m_q^{(k)}$ would not be 0 in the large $L$ limit: this subtraction prevents extra improvement coefficients to appear (see section 3 in \cite{LW1}) but, in practice, this should really matter only when working with renormalized quantities (while we deal with their bare counterparts).\\
\hspace*{0.7cm}Finally, it is possible to disentangle the effects of $\ \!c_{SW}^{(2)}$ and $\ \!c_{A}^{(2)}\ \!$ by means of $\ \!W$ and $\ \!W'$: in particular it turns out that, if the \emph{trivial background} (see footnote 4) is set, only $\ \!c_{A}^{(2)}$ has an effect at two-loop level. We start with this choice of the boundary gauge fields to fix this coefficient, afterwards $\ \!W$ and $\ \!W'$ will be changed to determine $c_{SW}^{(2)}\ \!$ thanks also to the by-then-known estimate of $\ \!c_{A}^{(2)}$.


\section{Numerical aspects}

Two more issues have still to be addressed about the present strategy, namely how configurations are generated and how the Wilson-Dirac operator is inverted to compute $f_A\ \!$ and $f_P$ eventually: to answer both, we must introduce some basics of NSPT$~\!$\footnote{$~\!$See \cite{FDR} and references therein for more details on this section in general.}.\\
\hspace*{0.7cm}Its core is given by the Langevin evolution equation that, for lattice gauge variables$~\!$\footnote{$~\!$As usual, fermion fields are integrated out so that only gauge degrees of freedom have to be eventually treated.}, reads

\begin{equation}
\frac{\partial}{\partial t}U_{\mu}(n,t)=-i\sum_{A}T^A\big[\nabla_{n\!,~\!\!\mu~\!\!,~\!\!A}S[U]+\eta^A_{\mu}(n,t)\big]U_{\mu}(n,t)\ ,
\end{equation}
\vspace*{-0.2cm}

\hspace*{-0.7cm}where $t$ is an extra degree of freedom (which can be thought as a \emph{stochastic time}), $S$ is the part of the lattice action depending on the $U$'s, $\eta$ is a Gaussian noise while $\nabla$ stands for the group derivative \newpage \hspace*{-0.7cm}defined as (index $``A"$ is summed over),

\begin{equation}
\mathcal{F}\big[e^{i~\!\!\alpha^A T^A}U_{\mu}(n),U'\big] = \mathcal{F}\big[U_{\mu}(n),U'\big] + \alpha^A\nabla_{n\!,~\!\!\mu~\!\!\!,~\!\!A}~\!\mathcal{F}[U_{\mu}(n),U'] + \ldots \ ,
\end{equation}
\vspace*{-0.2cm}

\hspace*{-0.7cm}where $T^A$ are the generators of the algebra and $\mathcal{F}$ is a generic scalar function of both the variable $U_{\mu}\!(n)$ and some more labelled $U'$ for short.\\ 
\hspace*{0.7cm}Given this setup, it can be shown that

\vspace*{-0.05cm}
\begin{equation}
Z^{-1}\!\!\int [DU]
 O[U]e^{-S[U]}=\lim_{t\rightarrow\infty}
 \frac{1}{t} \int_0^{\ \!\!t} \!\! {\rm d}\ \!t' \,
 \big\langle O[U_{\eta}(t')]\big\rangle_{\eta}\ ,
\end{equation}
\vspace*{-0.2cm}

\hspace*{-0.7cm}where $Z$ is the partition function and $O[U]$ a generic observable depending on the gauge fields.\\
\hspace*{0.7cm}Perturbation theory enters into play by \emph{formally} expanding each gauge degree of freedom in powers of $\beta_0^{-1}$ - defined as $\ \!\beta_0 = 2N_c/g_0^2\ \!$ being $\ \!N_c\ \!$ the number of colours - up to a given order $s$ as

\vspace*{-0.05cm}
\begin{equation}
U_{\mu}(n,t) = \mathbb{I} + \sum_{k=1}^s\beta_0^{-\frac{k}{2}}U_{\mu}^{(k)}(n,t)\ , 
\end{equation}
\vspace*{-0.15cm}

\hspace*{-0.7cm}and then plugging this Taylor series$~\!$\footnote{$~\!$Strictly speaking, Eq.(4.3) is valid only if the boundary gauge fields are set to the identity as in this first part of the study; once that a non-trivial \emph{background field} is introduced, the expansion would read $U_{\mu}(n,t) = exp[(C_k'-C_k)/T]\cdot\newline\cdot [\mathbb{I}+\sum_{k}\beta^{-\frac{k}{2}}U_{\mu}^{(k)}(n,t)$] $\ \!$-$\ \!$ consult section 6.2 in \cite{LW1} for the meaning of the first term in this product.} into Eq.(4.1): this results in a consistent \emph{hierarchical system of differential equations} which can be numerically integrated by discretizing the stochastic time as $t=n\tau\ \!$ with $\ \!n$ integer. In practice, the system starts from an arbitrary configuration and evolves by means of the solution of the discretized counterpart of Eq.(4.1): the desired observable is then obtained by averaging its measurements on its plateau - recall the limit in $t$ in Eq.(4.3)$~\!$\footnote{$~\!$This relation is true only for continuous $t$ so that simulations with different $\tau$ values have to be performed in order to extrapolate to $\tau\rightarrow 0$ afterwards.}.\\
\hspace*{0.7cm}As for the inverse of the fermionic operator, the entries needed to get $f_A$ and $f_P$ can be computed by means of the following perturbative formulae

\vspace*{-0.4cm}
\begin{eqnarray*}
\widetilde{M}^{-1^{(0)}}&=&\widetilde{M}^{(0)^{-1}}\ , \\
\widetilde{M}^{-1^{(1)}}&=&-\ \!\widetilde{M}^{(0)^{-1}}\widetilde{M}^{(1)}\widetilde{M}^{(0)^{-1}}\ , \\
\widetilde{M}^{-1^{(2)}}&=&-\ \!\widetilde{M}^{(0)^{-1}}\widetilde{M}^{(2)}\widetilde{M}^{(0)^{-1}} + \\
&&-\ \!\widetilde{M}^{(0)^{-1}}\widetilde{M}^{(1)}\widetilde{M}^{-1^{(1)}}\ ,\\
&\ldots&
\end{eqnarray*}
\vspace*{-0.4cm}

\hspace*{-0.7cm}where only the zeroth order of $\widetilde{M}$ has to be truly inverted: its expression for trivial $\ \!W$ and $W'$ can be found in section 3.1 of$\ \!$ \cite{LW2}.


\section{Preliminary results}

To test the correctness of the overall setup, we computed the one-loop contribution to $m_q$ without any counterterm subtraction for different choices of $\theta$ and $c_{SW}^{(0)}$$~\!$\footnote{$~\!$This is indeed the only $c_{SW}$ contribution that enters into play at this order with trivial $W$ and $W'$.} and compared the results \newpage \hspace*{-0.7cm}with the analytical values in Table 1.

\vspace*{0.15cm}
\begin{table}[h]
\begin{center}
\begin{tabular}{|c|c|c|c|}
\hline
$\theta$ & $c_{SW}^{(0)}=0.0$ & $c_{SW}^{(0)}=1.0$ & $c_{SW}^{(0)}=1.5$ \\
\hline
1.40 & 2.67621(4) & 1.67151(2) & 0.94999(1)    \\
\hline
1.00 & 2.63837(3) &  1.64808(1) & 0.93229(1)   \\
\hline
0.45 & 2.60727(3) & 1.62694(1) & 0.91948(1)   \\  
\hline
0.00 & 2.60571 & 1.62045 & 0.91067   \\  
\hline
\end{tabular}
\caption{Numerical results for $m_q^{(1)}$ on a $10^3*21$ lattice with $\ \!c_A^{(0)}=c_A^{(1)}=0$: the last line contains the infinite-volume results obtained from \cite{PAN}.}
\label{tab1}
\end{center}
\end{table}
\vspace*{-0.45cm}
It is reassuring that, when varying $c_{SW}^{(0)}$, outputs change accordingly: the still-existing gap is explained by recalling that finite-size effects are still present and that the analytical results correspond to $m_q^{(0)}=0$ while in our simulations $m_q^{(0)}\neq0$ due to the non-vanishing values of $\theta$ ($m_q^{(0)}$ approaches with decreasing $\theta$$~\!$\footnote{$~\!$An analytical expression for $m_q^{(0)}$ can be found in section 3 of \cite{LW2}.} the analytical infinite-volume values computed with $\theta=0.0$).


\section{Conclusions and acknowledgements}

According to the first, preliminary results, the outlined approach seems to be feasible: however, since different extrapolations (in $\tau$ and $L$) and interpolations (in $c_A^{(2)}$ and $c_{SW}^{(2)}$ when dealing with non-trivial $W$ and $W'$) are needed, extra care will have to be paid not to spoil accuracy.\\
\hspace*{0.7cm}We warmly thank \emph{LRZ} centre (Munich) and \emph{ECT$^*$} (Trento) for providing us with computer time on their clusters.\\
\hspace*{0.7cm}This work was supported by the DFG SFB/TR 55. 


\end{document}